\documentclass[aps,pra,reprint,twocolumn,superscriptaddress,showpacs,showkeys,superscriptaddress,longbibliography]{revtex4-2}
\usepackage{amssymb,amsfonts,amsmath,amstext,graphicx,hyperref}
\usepackage{tikz}
\usepackage{graphicx}
\usepackage[normalem]{ulem}
\hypersetup{
	colorlinks=true,
	linkcolor=blue,
	filecolor=magenta,      
	urlcolor=cyan,
}
\usepackage{xcolor}
\usepackage{amsthm}
\theoremstyle{plain}

\newtheorem*{theorem*}{Theorem}

\usepackage{braket}

\begin{document}

\title{Symmetries and Synchronization Blockade}

\author{Parvinder Solanki}
\affiliation{Department of Physics, Indian Institute of Technology-Bombay, Powai, Mumbai 400076, India}

\author{Faraz Mohd Mehdi}
\affiliation{Department of Physics, Indian Institute of Technology-Bombay, Powai, Mumbai 400076, India}
\affiliation{Department of Physics, University of Mumbai, Santacruz (East), Mumbai-400098, India}

\author{Michal Hajdu\v{s}ek}
\email{michal@sfc.wide.ad.jp}
\affiliation{Keio University Shonan Fujisawa Campus, 5322 Endo, Fujisawa, Kanagawa 252-0882, Japan}
\affiliation{Keio University Quantum Computing Center, 3-14-1 Hiyoshi, Kohoku, Yokohama, Kanagawa 223-8522, Japan}

\author{Sai Vinjanampathy}
\email{sai@phy.iitb.ac.in}
\affiliation{Department of Physics, Indian Institute of Technology-Bombay, Powai, Mumbai 400076, India}
\affiliation{Centre of Excellence in Quantum Information, Computation, Science and Technology, Indian Institute of Technology Bombay, Powai, Mumbai 400076, India}
\affiliation{Centre for Quantum Technologies, National University of Singapore, 3 Science Drive 2, 117543 Singapore, Singapore}

\date{\today}

\begin{abstract}
Synchronization blockade refers to an interferometric cancellation of quantum synchronization. In this manuscript, we show how the choice of synchronization measure and Hamiltonian symmetries affect the discussion of synchronization blockade. Using counting principles, we prove a general theorem that synchronization blockade cannot be observed in an $N-$level system when the coherent state used to define the diagonal limit-cycle state is in the full $SU(N)$ group. We present several illustrative examples of synchronization blockade in multi-level systems and prove that information-theoretic measures of synchronization can also observe synchronization blockade-like behavior by an appropriate choice of the set of limit cycle states.
\end{abstract} 
\maketitle


Motivated by experimental systems near their motional ground state \cite{zhang2012synchronization}, initial investigation of quantum synchronization was focused on infinite dimensional systems \cite{marquardt2006dynamical,lee2013quantum,manzano2013synchronization,walter2014quantum,walter2015quantum,ameri2015mutual,holmes2012synchronization,amitai2017synchronization,sonar2018squeezing,hush2015spin,xu2013synchronization,li2016quantum,li2017quantum,witthaut2017classical,bastidas2015quantum,heinrich2013collective,li2017properties,kato2019semiclassical,chia2020relaxation,kato2021enhancement}.
The theory of quantum synchronization has recently been extended to finite systems \cite{roulet2018synchronizing,roulet2018quantum,kwek2018no,koppenhofer2019optimal,tindall2020quantum,buca2022algebraic}, leading to its experimental observations in such systems \cite{laskar2020observation,krithika2022observation}.
These fundamental developments in our understanding of quantum synchronization lead to its application in other related fields such as quantum thermodynamics \cite{jaseem2020quantum,solanki2022role}, continuous time-translation symmetry breaking in boundary time crystals \cite{iemini2018boundary,hajduvsek2022seeding,krishna2022measurement}, and simulation of quantum dynamics on digital quantum computers \cite{koppenhofer2020quantum}.

The central question of fundamental importance is how synchronization in quantum systems differs when compared to classical systems.
Long-lived non-classical properties have been reported in strongly driven quantum systems in \cite{weiss2017quantum}, sometimes even surviving in the steady states \cite{sonar2018squeezing,lorch2016genuine}.
Other types of genuine quantum signatures include the presence of quantum discord \cite{mari2013measures} and entanglement \cite{roulet2018quantum} in the steady state.

An important phenomenon observed only in quantum systems is that of \textit{quantum synchronization blockade} \cite{lorch2017quantum,nigg2018observing}.
Initially, synchronization blockade due to \textit{frequency shifts} was investigated in systems of coupled infinite-dimensional oscillators in the deep quantum regime, where it was observed that detuned oscillators synchronize more strongly \cite{lorch2016genuine}.
Similar counterintuitive behavior was recently studied in single externally driven finite-dimensional systems \cite{koppenhofer2019optimal,tan2022blockade}.
Here, synchronization blockade was understood to set in as the result of \textit{destructive interference} leading to suppressed synchronization measures even for resonantly driven systems.
Despite its genuine quantum nature and immediate importance to understanding quantum synchronization, there is no systematic study of the general conditions under which synchronization blockade occurs.

In this manuscript, we develop a system-agnostic approach to quantum synchronization blockade based on the symmetries that a physical system obeys.
We derive a general condition for the existence of quantum synchronization blockade and show that $N$-level systems with fully unequal energy level spacings are not able to support quantum synchronization blockade.
On the other hand, synchronization blockade can be observed in systems with additional symmetries.
We show that understanding the conditions under which synchronization blockade may be observed is crucial in understanding the structure of limit cycle states.
Systems that do not support synchronization blockade must have fully diagonal limit cycle states, while systems with more symmetric energy level structure can have limit cycle states described by density matrices with coherences present.

Our manuscript is structured as follows.
In Section \ref{sec:sync_measure}, we discuss the phase-space based synchronization measure and derive its particularly useful form which we use throughout the manuscript.
We present our main result in Section \ref{sec:sync_blockade} where we derive a general condition for a system to support synchronization blockade.
We illustrate the application of our main result in Section \ref{sec:examples} by considering existing examples from the literature as well as a previously undiscussed four-level system.
In Section \ref{sec:comparison}, we discuss how our result fits within the context of information-theoretic measures of quantum synchronization.
We conclude with Section \ref{sec:discussion} and discuss how our approach can be applied to infinite-dimensional systems.

\section{Synchronization Measure}
\label{sec:sync_measure}

Synchronization measures in classical systems can be constructed using phase-space trajectories in a fairly straightforward fashion~\cite{pikovsky2003synchronization,strogatz2004sync} following the phase locking arguments.
For quantum systems on the other hand, the phase-space trajectories are not well defined and the notion of a phase-space is replaced by quasi-probability distributions~\cite{gerry2004introductory,schleich2011quantum} such as the  Wigner~\cite{wigner1932on}, the Husimi-Kano $Q$ ($Q$-function, for short) ~\cite{husimi1940some}, and the Glauber–Sudarshan $P$~\cite{sudarshan1963equivalence,glauber1963coherent} functions.
This has led to the introduction of phase-space-based measures of quantum synchronization, particularly for finite-level systems~\cite{roulet2018synchronizing,roulet2018quantum,jaseem2020quantum,krithika2022observation} where the genuine phenomenon of synchronization blockade~\cite{roulet2018synchronizing,koppenhofer2019optimal} has been discovered using such measures.
Other approaches not based on quasi-probability distributions exist as well \cite{li2017properties,jaseem2020generalized,galve2017quantum,mari2013measures}. our focus in the rest of this and the next section is phase space based measures, but we discuss the relationship of our results to quantum information theoretic measures in section IV.

We begin with a brief introduction to how these phase-space-based measures of quantum synchronization are constructed and how they herald the onset of quantum synchronization blockade.
We follow this with a general result proving that synchronization blockade can only exist for the systems following sub-algebra.

Consider a general $N$-level system whose unitary dynamics are described by transformations in the group $\mathcal{G}\equiv SU(N)$. Furthermore, let us  introduce the $Q$-function which can be defined over a coherent state
$\ket{\alpha}$ \cite{nemoto2000generalized,klauder1985coherent,radcliffe1971some,lee2015visualizing,mathur2002SU(N)} with group symmetry $\mathcal{H}$, with $\mathcal{H}\subseteq \mathcal{G}$. We will refer to the former as the \textit{full symmetry} of the problem whereas we refer to $\mathcal{H}$ as the symmetry of the bare Hamiltonian, since $\mathcal{H}$ is defined using the algebra of the bare Hamiltonian. The corresponding $Q$-function is defined as 
\begin{equation}
    Q(\vec\alpha) = \bra{\alpha}\rho\ket{\alpha} = \mathcal{N}\sum_{j,k}\alpha_j^*\rho_{jk}\alpha_k,\label{eq:qfun}
\end{equation}
where $\mathcal{N}$ is the normalization constant of the coherent state defined as $\mathcal{N}\int d^2 \alpha \ket{\alpha}\bra{\alpha}=\mathbb{I}$, and $\alpha_j=r_je^{-i \phi_j}$ is the coherent amplitude written in polar coordinates.
The $SU(N)$ coherent state is characterized by $2N-1$ phases \cite{nemoto2000generalized}.
Out of these, $N-1$ parameters $\vec{\theta}:=(\theta_1,\theta_2,\ldots,\theta_{N-1})$ are related to the populations of the state while the remaining $N$ parameters $\vec{\phi}:=(\phi_1,\phi_2,\ldots,\phi_{N})$ represent the free phases which also includes a global phase.
A synchronization measure accounting for the localization of free phases is constructed by integrating out the population degrees of freedom from the $Q$-function \cite{roulet2018quantum,roulet2018synchronizing,jaseem2020quantum,krithika2022observation},
\begin{align}
    S(\vec{\phi})=\int d\Omega_\theta Q(\vec{\theta},\vec{\phi})-C.\label{eq:sync_def}
\end{align}
Here $d\Omega=d\Omega_\theta d\Omega_\phi$ is the Haar measure \cite{tilma2002parametrization} for the group $\mathcal{H}$, while $d\Omega_\theta,$ $d\Omega_\phi$ are the contributions of population degrees of freedom and phases to the Haar measure respectively, and $C$ is the equiprobable distribution of the free phases.
This means that the synchronization measure in Eq.~(\ref{eq:sync_def}) quantifies the degree of synchronization as the deviation from a uniform phase distribution.

For a general $N$-level system, let us write the complex components of the density matrix as $\rho_{jk}=R_{jk}e^{-i\chi_{jk}}$.
Using Eq.~(\ref{eq:qfun}) and Eq.~(\ref{eq:sync_def}),  the synchronization measure can be expressed as follows,
\begin{align}
    S(\vec{\phi})=\mathcal{N}\sum_{j,k=1}^{N}R_{jk}z_{jk}(\cos\xi_{jk}+i \sin\xi_{jk})-\dfrac{1}{(2\pi)^{N-1}},\label{eq:sync_measure_gen}
\end{align}
where $\xi_{jk}=\phi_j-\phi_k-\chi_{jk}$. Integration over the population parameters is given by $z_{jk}=\int d\Omega_\theta r_{j}r_{k}$ and $(1/2\pi)^{N-1}$ represents the equiprobable distribution of the $N-1$ free phases excluding the global phase.

The term $\sum_{j,k}R_{jk}z_{jk}\sin\xi_{jk}$ in Eq.~(\ref{eq:sync_measure_gen}) vanishes since $R_{jk}$ and $z_{jk}$ are symmetric but the term $\sin\xi_{jk}$ is anti-symmetric under  $j\leftrightarrow k$.
The summation over diagonal terms $\sum_{j=k} R_{jk}z_{jk}\cos\xi_{jk}$ is equal to the equiprobable distribution of free phases (see Appendix \ref{appendix1}) which cancels the last term of Eq.~(\ref{eq:sync_measure_gen}).
Therefore the synchronization measure given by Eq.~(\ref{eq:sync_measure_gen}) reduces to
\begin{align}
    S(\vec{\phi})=2\mathcal{N}\sum_{j<k}^{N} R_{jk}z_{jk}\cos\xi_{jk},
    \label{eq:final_gen_sync_meas}
\end{align}
where we have replaced the full sum with twice the sum for $j<k$ given that $\cos\xi_{jk}$ is symmetric under $j \leftrightarrow k$.
In the following sections, we will be using Eq.~(\ref{eq:final_gen_sync_meas}) to study the synchronization blockade in sub-group systems. A similar derivation was presented in \cite{tan2022blockade}.

Apart from phase-space-based measures, a quantum information theoretic measure was introduced recently in \cite{jaseem2020generalized}.
Such measures use a system-independent approach to quantify the amount of synchronization present in a given quantum system and provide a generalized synchronization measure.
The basic idea of such measures is to quantify the distance of a given state from the nearest possible limit cycle state which lacks synchronization.
Therefore, we define a set of limit cycle states over which a distance measure can be optimized to calculate the amount of synchronization present in a system.
For an externally driven unipartite system, such measures reduce to the $l_1$ norm of coherence~\cite{jaseem2020generalized} given by $\sum_{j\neq k}\vert\rho_{jk}\vert$.
Therefore a system having a finite amount of coherence will show a non-zero value for the synchronization measure.

Interestingly, in \cite{koppenhofer2019optimal} the authors showed that for blockade conditions, phase-space-based measure vanishes even in the presence of coherences for an externally driven spin-$1$ system. We note that an equivalent $l_1$ norm based measure of synchronization does not register blockade for such a system. We will hence try to understand what the relationship between blockade and coherences are in detail in the upcoming sections.
Since the formal theory of synchronization blockade for the spin-$1$ system has been introduced using phase-space based measure, we first investigate the blockade conditions for a general $SU(N)$ system using this measure in the next section.




\section{Synchronization Blockade}
\label{sec:sync_blockade}

Synchronization blockade is understood as the destructive interference between the coherences in a system capable of showing synchronous behavior.
This results in the synchronization measure being zero even in the presence of coherences irrespective of the values of free phases.
The exact role of system symmetries leading to destructive interference of coherences has not been discussed so far in the literature.
There exist no general criteria to establish whether a given system can exhibit synchronization blockade or not.
In this section, we study all these issues systematically by using group theoretic methods.  
We begin by proving the following theorem relating the subgroup $\mathcal{H}$ and the synchronization blockade:  


\begin{theorem*}
Let $\mathcal{H}\subseteq\mathcal{G}$ be the subgroup over which the coherent states are defined such that the undriven limit cycle states are diagonal as defined by Eq.(1). Synchronization blockade is not permitted if $\mathcal{H}=\mathcal{G}$. 


\end{theorem*}

\emph{Proof:} From Eq.~(\ref{eq:final_gen_sync_meas}), the blockade condition is given by,
\begin{align}
    S(\vec{\phi})=2\mathcal{N}\sum_{j<k}a_{jk}\cos\xi_{jk}=0,
\label{eq:blockade_cond}
\end{align}
where, $a_{jk}=R_{jk}z_{jk}$ and $z_{jk}\neq 0$ since the coherent states are not orthogonal.
When $\mathcal{H}=\mathcal{G}$, all the free phases $\{\phi_1,\phi_2,\ldots,\phi_N\}$ are independent of each other.
Therefore, the total number of independent $\cos \xi_{jk}$'s is $^NC_2$, which is equal to the total number of terms in Eq.~(\ref{eq:blockade_cond}).
In such case, the only possible solution of Eq.~(\ref{eq:blockade_cond}) is when all the coefficients $a_{jk}$ are 
individually zero (see Appendix \ref{appendix2}).
Therefore all the coherences are also zero and we are left with a diagonal density matrix which corresponds to a limit cycle state \cite{jaseem2020quantum,jaseem2020generalized}.
This concludes the proof that if the coherent state for an $N-$level system is constructed in the group $SU(N)$, then the interferometric cancellations will not happen and thus synchronization blockade does not exist. 


Consider the group generated by the dynamical lie algebra \cite{qcontrol} of the bare Hamiltonian to be a proper subgroup $\mathcal{H}\subset \mathcal{G}$; say $\mathcal{H} \equiv SU(M)$ with $M<N$.
Then the terms in Eq.~(\ref{eq:blockade_cond}) will generally not be linearly independent.
This results from the fact that the number of free phases, $M$, for any given proper subgroup is less than the number of free phases required to address the full group $\mathcal{G}$.
In other words, the total number of independent $\cos\xi_{jk}$'s will be given by $^MC_2$, which is less than the total number of terms, $^NC_2$, in Eq.~(\ref{eq:blockade_cond}).
Therefore there \textit{might be} terms in the linear combination which are not independent.
Now we can have a non-trivial solution of this linear combination such that the non-zero density matrix terms cancel each other to provide synchronization blockade.
In particular, the blockade condition is satisfied if $|\phi_j-\phi_k|$ is the same for at least two $a_{jk}$'s.
The terms $a_{jk}$ and $\chi_{jk}$ depend on the density matrix elements which can be tuned by varying the system parameters to satisfy the blockade condition.
This results in the synchronization measure being zero even for non-zero $a_{jk}$'s and hence one might observe synchronization blockade if the coherent state is constructed in the proper subgroup of $\mathcal{G}$. We note here that though a subgroup construction is necessary for observing synchronization blockade, it is not sufficient. 

\begin{figure}[tp!]
    \centering
    \includegraphics[width=\columnwidth]{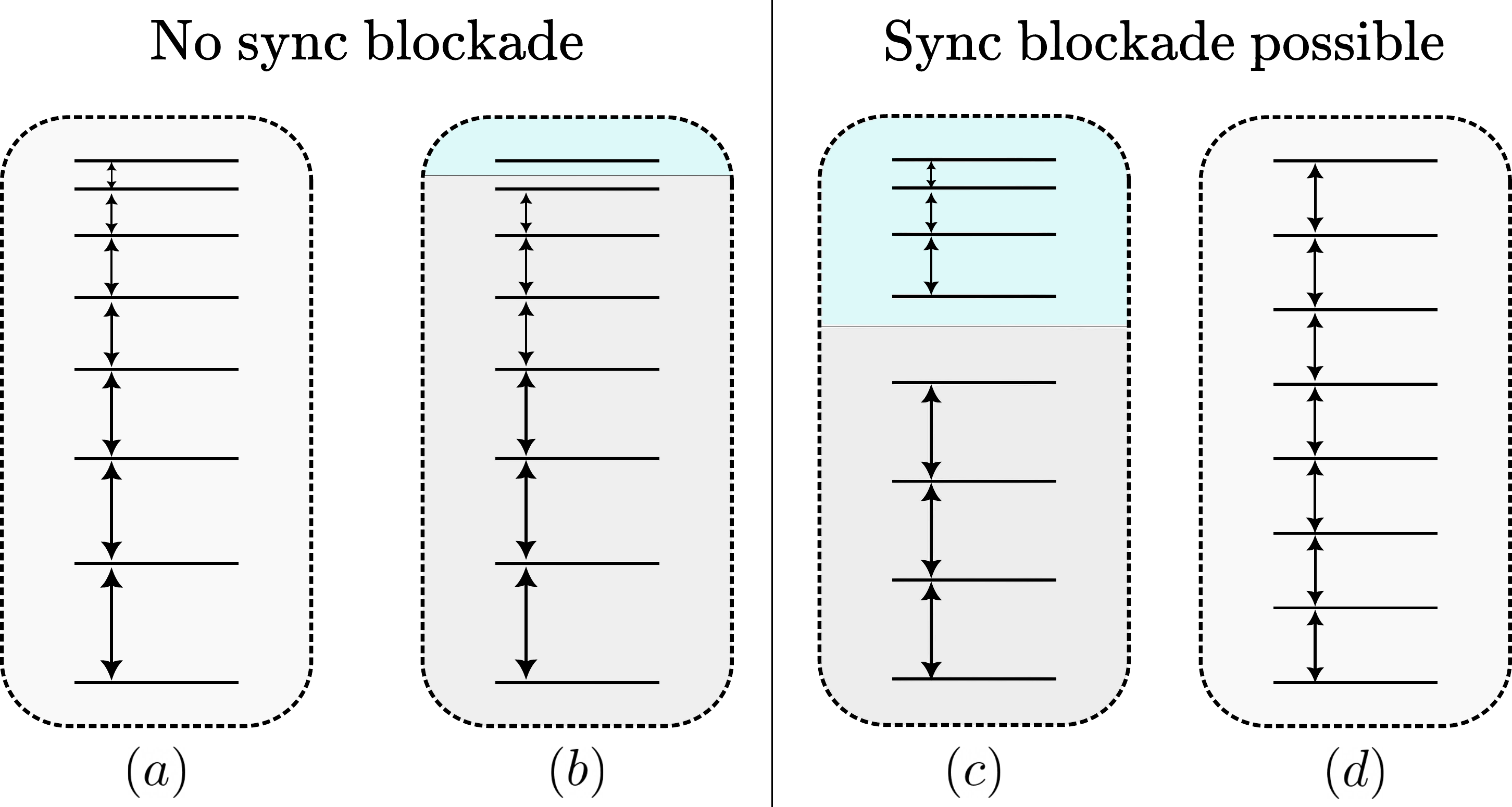}
    \caption{Here is an example of different subgroups of a general $8$-level system (sub-figure (a)).
    Sub-figures (a) and (d) represent the fully connected $8$ level system following $\mathfrak{su}(8)$ and $\mathfrak{su}(2)$ algebra, respectively. Whereas (b) represents the trivial $U(1)\oplus SU(7)$ subgroup leaving one level disconnected from the rest of the levels and (c) represents a more complicated sub-group $SU(4)\oplus SU(2)$ of $SU(8)$. Synchronization blockade can be observed only if the fully connected $N$-level system follows a sub-group algebra $SU(M)$ with $M<N$ (sub-figure (d)) or one/all of the disjoint $d-$ dimensional block diagonal structures follows sub-group algebra $SU(l)$ with $l<d$ (sub-figure (c)). 
    }

    \label{fig:su8_subgroups}
\end{figure}

The above theorem can be well understood from 
Fig.~\ref{fig:su8_subgroups}(a) which represents an unequally spaced $N$ level system where all levels are connected via nearest neighbor transitions. 
In the appendix \ref{appendix3}, we prove that such a systems dynamics is described by the algebra $\mathfrak{su}(N)$ 
($\mathcal{H}=\mathcal{G}=SU(N)$) and hence cannot have a synchronization blockade following from the theorem stated above. 
If fully connected systems follow a sub-algebra of $\mathfrak{su}(N)$ as shown in  Fig.~\ref{fig:su8_subgroups}(d), it can exhibit synchronization blockade. 

On the other hand, we can have a disconnected system as shown in Fig.~\ref{fig:su8_subgroups}(b) and Fig.~\ref{fig:su8_subgroups}(c).
The total Hilbert space $\mathbf{H}(N)$ of such systems can be written as a direct sum of $k-$Hilbert spaces $\oplus_k \mathbf{H}(N_k)$ such that $N=\sum_{k} N_k$ where $N_k$'s gives the Hilbert space dimension of disconnected subspaces.
In such a case, the given theorem applies to all the Hilbert spaces individually.
Total synchronization measure for such cases will be a linear combination of the synchronization measure of disconnected sub-spaces, $S(\vec{\phi})=\sum_k S_{N_k}(\vec{\phi}_{N_k})$. Therefore synchronization blockade of a given sub-system $\mathcal{H}_k$ is only reflected in the corresponding synchronization measure $S_{N_k}(\vec{\phi}_{N_k})=0$ and manifests itself as a minimum of total synchronization measure $S(\vec{\phi})$.
In Fig.~\ref{fig:su8_subgroups}(b), one of the levels is disconnected from the rest and does not interact with other levels.
Hilbert space for this system is $\mathbf{H}(1)\oplus \mathbf{H}(7)$ and follows $\mathfrak{u}(1)\oplus \mathfrak{su}(7)$ algebra.
This can be understood from the appendix \ref{appendix3} where we prove that a system of a general $M$ connected levels with unequal energy gaps will follow $\mathfrak{su}(M)$ algebra.
As an example, in the individual Hilbert spaces, $\mathcal{H}_{N_i}=\mathcal{G}_{N_i}$, and hence we do not observe synchronization blockade.
In Fig.~\ref{fig:su8_subgroups}(c), the upper four levels are unequally spaced and follow $\mathfrak{su}(4)$ algebra while the lower four levels are equally spaced following the algebra $\mathfrak{su}(2)$.
The dynamics of this whole system is hence well described by $\mathfrak{su}(4)\oplus\mathfrak{su}(2)$ algebra in a $\mathbf{H}(4)\oplus \mathbf{H}(4)$ Hilbert space such that $N_{1,2}=4$. 
Synchronization blockade can only be observed for the lower group of equally spaced four levels following  $\mathcal{H}_{N_2}=SU(2)\subset \mathcal{G}_{N_2}=SU(4)$ and hence will be reflected as a minima of total synchronization measure.
The minima of full synchronization measure will be zero only when $\mathcal{H}_{N_1}$ exists in the limit cycle state and $\mathcal{H}_{N_2}$ exhibits synchronization blockade. We discuss several examples in the next section to illustrate the use of our theorem.


\section{Examples}
\label{sec:examples}

There exist several examples of open quantum systems exhibiting synchronization blockade under certain specific conditions \cite{roulet2018quantum,koppenhofer2019optimal,laskar2020observation,koppenhofer2020quantum}.
In this section, we discuss such examples to show that synchronization blockade exists only if the {coherent state is constructed in a proper subgroup of $\mathcal{G}.$}
We begin by discussing an existing example of synchronization blockade for a spin-$1$ system \cite{roulet2018synchronizing}.
We then introduce a new example of a spin-3/2 system exhibiting the synchronization blockade.
Finally, we show that synchronization blockade cannot exist for a three-level system following the {$\mathfrak{su}(3)$} algebra.

\subsection{$\mathcal{H}\equiv SU(2)$,  $\mathcal{G}\equiv SU(3)$}
\label{sec:su2su3}
Since a qubit does not have an observable free phase and it lacks a valid limit cycle state \cite{roulet2018synchronizing,solanki2022role}.
The next smallest system to study synchronization \cite{roulet2018synchronizing} and hence synchronization blockade \cite{koppenhofer2019optimal} is a spin-$1$ system.
The spin-$1$ system can be described by $\mathcal{H}\equiv SU(2)$ group which is a {proper} subset of $\mathcal{G}\equiv SU(3)$ group, describing a general three-level system.
Therefore it is a perfect candidate to study synchronization blockade.
Here we discuss the synchronization blockade for the spin-$1$ system considered in \cite{roulet2018synchronizing}.
The master equation describing the dynamics of the driven-dissipated spin-$1$ system in the rotating frame of the drive is given by 
\begin{equation}
    \Dot{\rho}=-i[\Delta S_z+\varepsilon S_y, \rho]+\gamma_g\mathcal{D}[S_+ S_z]\rho+\gamma_d\mathcal{D}[S_- S_z]\rho,
    \label{eq:su2bruder}
\end{equation}

where $S_{x,y,z}$ represent the usual Spin-$1$ Pauli-matrices \cite{koppenhofer2019optimal}.
Interaction with the environment is modelled by the Lindblad dissipators of the form $\mathcal{D}[O]=O\rho O^\dagger -\frac{1}{2}\{O^\dagger O, \rho\}$ where $O$ represents the jump operators.
Furthermore,  $\Delta = \omega_d - \omega_0$ is the detuning between the drive frequency ($\omega_d$) and the natural frequency ($\omega_0$) of the spin-$1$ system, $\gamma_g$, and $\gamma_d$ are dissipation rates for the gain and damping modelled by jump operators $S_+ S_z$ and $S_- S_z$, respectively.
In the absence of the drive, the system settles down to the diagonal steady state where {the} population of extremal levels dissipates to the middle one.
The $Q$-function defined by the $SU(2)$ coherent state ,
\begin{align}
    \ket{\alpha}=\begin{pmatrix} e^{i\phi}\cos^2(\theta/2)\\ \sin(\theta)/\sqrt{2}\\ e^{-i\phi} \sin^2(\theta/2)\end{pmatrix},
\end{align}
shows no localization of the free phase in the steady state.
The semi-classical signal given by $\varepsilon S_y$ drives both transitions $\ket{1}\leftrightarrow\ket{2}$ and $\ket{3}\leftrightarrow\ket{2}$.
In the presence of this external drive, the system shows phase localization in $\phi$ and synchronization measure $S(\phi)$ registers a non-zero value.
Since the drive strength $\varepsilon$ is perturbatively small and no dissipator is acting between the extremal levels, we can consider $\rho_{13}=R_{13}e^{-i\chi_{13}}=0$.
The synchronization measure defined in Eq.~(\ref{eq:sync_measure_gen}) reduces to the following form for the given system,
\begin{align}
    S(\vec{\phi}) & = 2\mathcal{N} \left[ z_{12}R_{12}\cos(\xi_{12})+z_{23}R_{23}\cos(\xi_{23}) \right],\nonumber\\
    & = \dfrac{\pi \mathcal{N}}{2\sqrt{2}}\left[ R_{12}\cos(\phi+\chi_{12})+R_{23}\cos(\phi+\chi_{23}) \right].
    \label{eq:su2su3}
\end{align}
In order to observe synchronization blockade, $S(\phi)$ needs to vanish which leads to the condition $R_{12}/R_{23}=-\cos(\phi+\chi_{23})/\cos(\phi+\chi_{12})$.
Now, if we take $R_{12}=R_{23}$ and $\chi_{12}=\chi_{23}\pm\pi$ we obtain $\rho_{12}=-\rho_{23}$. 
This leads to destructive interference between the coherences resulting in synchronization blockade.
These conditions can be easily satisfied by tuning the parameters in the master equation appropriately.
Blockade condition $\rho_{12}=-\rho_{23}$ for the spin-$1$ system described by Eq.~(\ref{eq:su2bruder}) is satisfied for $\gamma_g=\gamma_d$ which has been studied in \cite{koppenhofer2019optimal}.
This example verifies that for the {coherent state constructed in a proper} sub-group {of SU(N)}, the synchronization measure reduces to the linear combination of coherences which can be set to zero to observe synchronization blockade, independent of the free phases.

\begin{figure}[t!]
    \centering
    \includegraphics[width=\columnwidth]{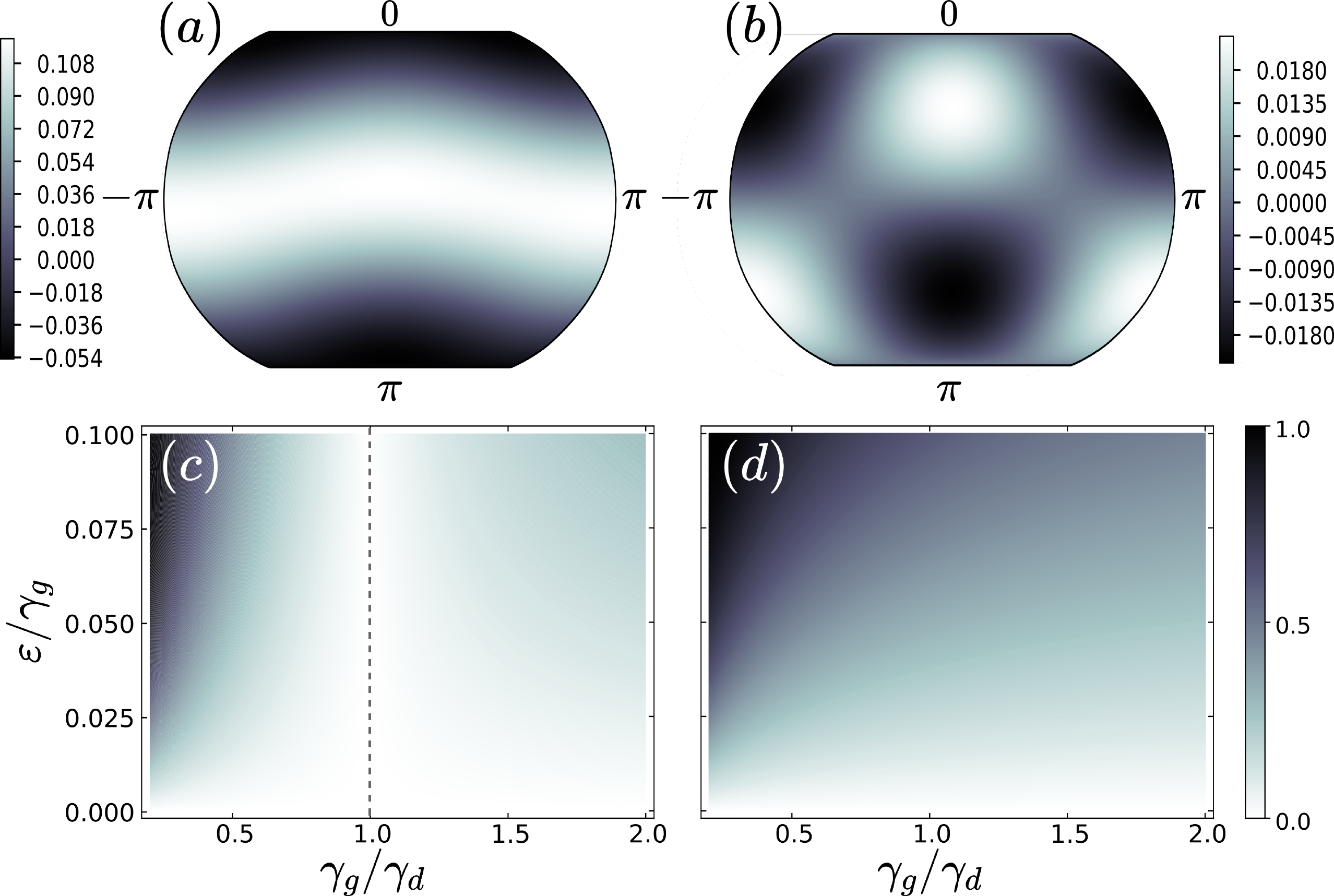}
     \caption{Spin $Q$-functions and synchronization measure for parametric variation. Sub-figure (a) represents the $Q$-function of the steady state $\rho_{ss}$ of the spin-$1$ system described by Eq.~(\ref{eq:su2bruder}) for parameter values $\gamma_g=\gamma_d=0.1,\varepsilon=0.1\gamma_g$ and $\Delta=0$. The contribution of the off-diagonal elements of $\rho_{ss}$ to the $Q$-function is given by sub-figure (b). It depicts the phase localization of $\phi$ which is not clear from (a). Phase space-based synchronization measure $S(\phi)$ represented in (c) vanishes for $\gamma_d=\gamma_g$. While the distance-like synchronization measure given by $l_1$ norm for a diagonal limit cycle state is non-zero for $\gamma_g=\gamma_d$ as shown in (d). Both measures are re-scaled to their maximum values for the parameters values of $\Delta=0$ and $\gamma_g=0.1$.}
    \label{fig:comparison}
\end{figure}

To further understand the role of coherences in synchronization blockade, we investigate the $Q$-function of the spin-$1$ system considered in this section.
Figure \ref{fig:comparison}(a) represents the $Q$-function for the steady state of Eq.~(\ref{eq:su2bruder}) and  shows the equiprobable distribution of phase $\phi$.
Whereas Fig. \ref{fig:comparison}(b) illustrates the contribution of only the off-diagonal {density matrix} elements to the $Q$-function.
We observe that for the blockade condition, the contribution of the coherences to the $Q$-function is much smaller than the diagonal elements.
Therefore the variation due to coherences is not observed in the $Q$-function represented in Fig. \ref{fig:comparison}(a).
Interestingly, the contribution of the off-diagonal elements to the $Q$-function is localized in phase $\phi$ as shown in Fig. \ref{fig:comparison}(b).
Also, the magnitude of the $Q$-function exhibits equal and opposite  distribution over the values of $\theta$ for $\gamma_g=\gamma_d$.
Due to this specific distribution, the integration of $Q$-function {over population degrees of freedom} vanishes and results in the observation of synchronization blockade.
Therefore the synchronization measure $S(\phi)$ vanishes for $\gamma_d=\gamma_g$ as shown in Fig.~\ref{fig:comparison}(c). 

\subsection{$\mathcal{H}\equiv SU(2)$, 
$\mathcal{G}\equiv SU(4)$}
\begin{figure*}[htp!]
    \centering
    \includegraphics[width=0.9\textwidth]{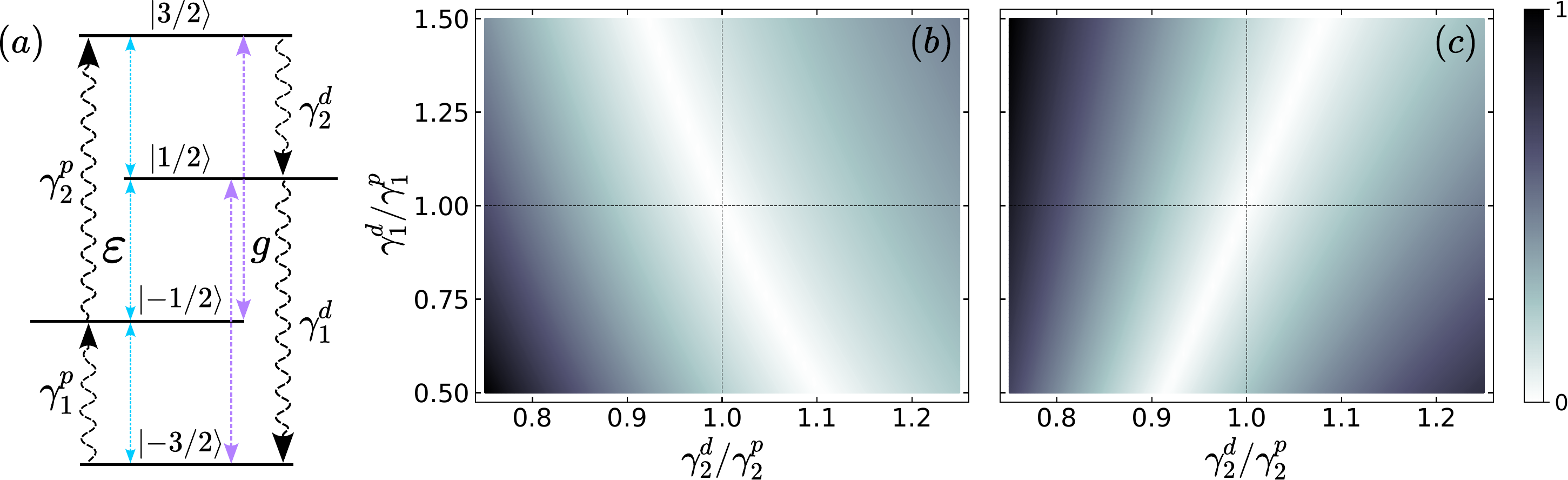}
    \caption{
    (a) Schematic representation of an equally spaced four-level system coupled to four different baths (given by curly dotted lines along with their dissipation rates) as described by Eq.~(\ref{eq:4levelmaster}). Drive $V_1$ and $V_2$ are represented by blue and purple transition lines, respectively. Subfigures (b) and (c) represent the parameter regime over which the synchronization measure given by Eq.~(\ref{eq:su2su4}) vanishes and displays blockade for drive $V_1 (\varepsilon=0.1\gamma^p_1,g=0)$ and $V_2 (\varepsilon=0,g=0.1\gamma^p_1)$, respectively. Other parameter values are $\Delta=0,\gamma^p_1=0.1$ and $\gamma^p_2=1$.}
    \label{fig:su2su4}
\end{figure*}

We now consider the 
$SU(2)$ {coherent state for a 4-level system,} 
depicted by Fig.~\ref{fig:su2su4}(a).  
The coherent state construction of this system is similar to that of the spin-$1$ system.
Therefore after rotating the extremal state of the spin-3/2 system using the four-dimensional representation of $\mathcal{H}$, the coherent state is given by
\begin{align}
    \ket{\alpha}=\begin{pmatrix} e^{-3i\phi/2}\cos^3(\theta/2)\\ \sqrt{3}e^{-i \phi/2}\cos^2(\theta/2)\sin(\theta/2)\\\sqrt{3}e^{i \phi/2}\cos(\theta/2) \sin^2(\theta/2)\\ e^{3i\phi/2}\sin^3(\theta/2)\end{pmatrix}.
\end{align}
 For this coherent state of the spin-3/2 system, synchronization measure defined in Eq.~(\ref{eq:sync_measure_gen}) simplifies to
\begin{eqnarray}
    S(\vec{\phi}) & = & 2\mathcal{N}\Big[\dfrac{5\pi\sqrt{3}}{64}\{ R_{12}\cos(\phi+\chi_{12})
    \nonumber\\
    &+& \frac{3\sqrt{3}}{5}R_{23}\cos(\phi+\chi_{23}) +R_{34}\cos(\phi+\chi_{34}) \} \nonumber\\ &+&\dfrac{\sqrt{3}}{6}\{ R_{13}\cos(2\phi+\chi_{13})
    + R_{24}\cos(2\phi+\chi_{24})\}
     \nonumber\\ &+& \dfrac{3\pi}{64}R_{14}\{\cos(3\phi+\chi_{14})\}\Big].
    \label{eq:su2su4}
\end{eqnarray}
It can be observed that the synchronization measure has three sets of independent terms (given in $\{ \cdot \}$) which can vanish independently to satisfy the synchronization blockade condition.

We now construct a specific model of a four-level system with equally spaced energy levels as shown in Fig.~\ref{fig:su2su4}(a) which is capable of displaying synchronization blockade.
The dynamics of this system is given by the following master equation,
\begin{eqnarray}
    \Dot{\rho}&=&-i[H_0+\varepsilon V_1+gV_2,\rho]+\gamma^p_1\mathcal{D}[\sigma_{-\frac{1}{2},-\frac{3}{2}}]\rho+\nonumber\\ & &\gamma^p_2\mathcal{D}[\sigma_{\frac{3}{2},-\frac{1}{2}}]\rho+ \gamma^d_1\mathcal{D}[\sigma_{\frac{1}{2},\frac{3}{2}}]\rho+\gamma^d_2\mathcal{D}[\sigma_{-\frac{3}{2},\frac{1}{2}}]\rho ,\label{eq:4levelmaster}
\end{eqnarray}
where $H_0=\Delta S_z$, $\sigma_{j,k}=\vert j \rangle \langle k \vert$ and $\gamma^{m}_n$ are dissipation rates such that $m=\{p,d\}$ and $n=\{1,2\}$.
The terms $V_1$ and $V_2$ correspond to two external drives $S_x$ (blue transition lines in Fig.~\ref{fig:su2su4}(a)) and $S_x^2$ (purple transition lines in Fig.~\ref{fig:su2su4}(a)), respectively.
The drives are always resonant with their respective transitions which lead to $\Delta=0$.
We will now discuss the synchronization blockade conditions for drives $V_1$ and $V_2$.

Let us first consider the blockade condition for drive $\varepsilon V_1$ alone by setting $g=0$.
For perturbative values of  $\varepsilon$, only the coherence between two adjacent levels will contribute and the rest will vanish, \textit{i.e.,} $R_{13}=R_{14}=R_{24}=0$. This reduces Eq.~(\ref{eq:4levelmaster}) to the following blockade condition,
    \begin{align}
        5\sqrt{3}\left[R_{12}\cos(\phi+\right.&\left.\chi_{12})+R_{34}\cos(\phi+\chi_{34})\right]\nonumber\\
        &+9R_{23}\cos(\phi+\chi_{23})=0.
        \label{eq:blockade_Sx}
    \end{align}
One particular way of simplifying this blockade condition is when the arguments of all non-zero coherences between the adjacent levels follow the relation $\chi_{34}=\chi_{12}$ and $\chi_{23}=\chi_{12}\pm\pi$.
This leads to the blockade condition $5\sqrt{3}(R_{12}+R_{34})=9R_{23}$.
The parameter regime for which the synchronization measure vanishes and results in synchronization blockade is shown in Fig.~\ref{fig:su2su4}(b).
Naturally, there are other possible solutions satisfying Eq.~(\ref{eq:blockade_Sx}), which lead to destructive interference between the coherences of the steady state resulting in synchronization blockade.
    
We now consider the effect of the drive $gV_2$ only by setting $\varepsilon=0$.
Under the effect of a perturbative drive $V_2$, coherences between adjacent and extremal levels will vanish, $R_{12}=R_{23}=R_{34}=R_{14}=0$.
This leads to the following blockade condition,
\begin{align}
    \dfrac{\sqrt{3}}{6}\left[R_{13}\cos(2\phi+\chi_{13})+R_{24}\cos(2\phi+\chi_{24})\right]=0,
\end{align}
which is similar to the blockade condition of the spin-$1$ system discussed in Sec.~\ref{sec:su2su3}.
Therefore, the two coherences need to be equal in magnitude ($R_{13}=R_{24}$) but have $\pi$ phase difference ($\chi_{13}=\chi_{24}\pm\pi$) such that $\rho_{13}=-\rho_{24}$.
The parameter regime for which synchronization blockade appears for drive $V_2$ is shown in Fig.~\ref{fig:su2su4}(c).

Recently, synchronization blockade for higher spin systems following {$\mathfrak{su}(2)$} algebra was discussed in \cite{tan2022blockade}.
The authors considered a general scheme of dissipators that defines the symmetry of the underlying limit cycle, resulting in different synchronization blockade conditions.
We note that the synchronization blockade discussed in \cite{tan2022blockade} can be well-understood within the paradigm of our theorem.
For instance, the authors consider an example of a spin-3/2 system coupled with distinct dissipators resulting in different blockade conditions.
It can be easily verified that Eq.~(\ref{eq:su2su4}) vanishes under all such blockade conditions irrespective of the underlying dissipator scheme of the spin-3/2 system.
This reinforces our main result that synchronization blockade can only emerge in the systems following sub-group algebra and can be explored via our approach.

\subsection{No blockade for 
$\mathcal{H}\equiv\mathcal{G}\equiv SU(3)$ group}
We now consider an unequally spaced three-level system 
{in which the coherent state is constructed under the $SU(3)$ group, i.e., $\mathcal{H}\equiv SU(3)\equiv\mathcal{G}$} as discussed in \cite{jaseem2020quantum}.
The authors consider a driven three-level system coupled with two separate baths, kept at different temperatures.
The evolution of such a system in the rotating frame of the external drive with frequency $\omega_d$ is given by,
\begin{align}
    \Dot{\rho}=-i[H_0+\varepsilon V,\rho]+\mathcal{L}_h[\rho]+\mathcal{L}_c[\rho],
\end{align}
where $H_0=\Delta \sigma_{33}$ such that $\Delta=\omega_3-\omega_2-\omega_d$ with $\omega_i$ being the energy of $i_{th}$ level.
The drive $V=\sigma_{23}+\sigma_{32}$ couples the upper two levels.
Lindblad superoperators $\mathcal{L}_h[\rho]$ and $\mathcal{L}_c[\rho]$ model the hot and cold baths coupled to extremal and lower two levels, respectively.
For such a general three-level system, the coherent state is given as
\begin{align}
   \ket{\alpha}= \begin{pmatrix}\cos\theta_1\\ e^{i\phi_1}\cos\theta_2 \sin\theta_1\\ e^{i\phi_2}\sin\theta_2 \sin\theta_1 \end{pmatrix}.
\end{align}
We note that due to the structure of 
{$\mathfrak{su}(3)$} algebra the system possesses two free phases represented by $\phi_1$ and $\phi_2$. The rest of the two phases, $\theta_1$ and $\theta_2$, are related to the populations of the state.
The synchronization measure needed to study its synchronization properties, given in Eq.~(\ref{eq:sync_measure_gen}) reduces to, 
\begin{align}
    S(\vec{\phi})&=\dfrac{1}{8\pi}[R_{12}\cos(\phi_1-\chi_{12})+R_{13}\cos(\phi_2-\chi_{13})\nonumber\\
    & + R_{23}\cos(\phi_2-\phi_1-\chi_{23})].
    \label{eq:su3}
\end{align}

In the above equation, all the terms are linearly independent and hence synchronization measure in Eq.~(\ref{eq:su3}) vanishes only when all the coherence terms are zero.
Therefore synchronization blockade is possible only when all the coherence terms $R_{ij}$ are individually zero.
This verifies our theorem that synchronization blockade cannot be established in an $N-$level system where the coherent state is constructed in $SU(N)$.

\section{Comparison of various Synchronization Measures}
\label{sec:comparison}

In this section, we explore the connection between synchronization blockade within the context of phase-space based measure of synchronization of Eq.~(\ref{eq:sync_def}) and information-theoretic measures of synchronization introduced in \cite{jaseem2020generalized}.
These generalized measures quantify synchronization as the distance between the steady state of the evolution $\rho$ and the nearest limit cycle,
\begin{equation}
    \Omega (\rho) \equiv \min_{\sigma \in \Sigma} \mathfrak{D} (\rho, \sigma).
    \label{eq:gen_measure}
\end{equation}
Distance measures discussed for the measure $\mathfrak{D}(\rho,\sigma)$ include the relative entropy $S(\rho||\sigma) = \text{Tr}\left\{ \rho\log\rho_{ss} - \rho\log\sigma \right\}$ and the trace distance $||\rho-\sigma||_1 = \text{Tr} \{ \sqrt{(\rho-\sigma)^{\dagger}(\rho-\sigma)}\}$.
Regardless of the choice for the distance measure, the central object is the set of all limit cycle states $\Sigma$.
This presents an immediate challenge of determining the appropriate structure of the limit cycle over which the minimization needs to be carried out.

Our theorem presents an important tool for identifying the correct structure of such limit cycle states.
In the case of an $N$-level quantum system with fully unequal energy spacings, the structure of a limit cycle is rather straightforward.
Such a system follows the group $\mathcal{G}$
 and therefore does not permit synchronization blockade.
This in turn implies that the limit cycle states must be diagonal, which greatly simplifies the minimization in Eq.~(\ref{eq:gen_measure}).
Such a scenario arises naturally in the context of quantum thermodynamics \cite{jaseem2020quantum}.
Further simplification of evaluation of distance-based measures of quantum synchronization in Eq.~(\ref{eq:gen_measure}) is that due to the diagonal structure of the limit cycle states for  $\mathcal{G}$, one can use the $l_1$-norm as a suitable measure.
In that case we have $\Omega_{l_1}(\rho) = \sum_{j\neq k} |\rho_{jk}|$.

The situation changes when the $N$-level physical system follows a proper subgroup $\mathcal{H}$ of the group $\mathcal{G}.$  
Such a system is capable of supporting the synchronization blockade where the phase-space based measure of quantum synchronization vanishes even when the steady state density matrix contains finite coherences.
The information-theoretic measure of quantum synchronization in Eq.~(\ref{eq:gen_measure}) must reflect this as well which means the limit cycle set of states must be expanded to include non-diagonal density matrices.
Our Theorem provides a useful parameterization of the permissible non-diagonal limit cycle states as all of them must satisfy the synchronization blockade condition in Eq.~(\ref{eq:blockade_cond}).

To illustrate this point we plot the synchronization measure $S(\phi)$ for an externally driven spin-$1$ system evolving according to Eq.~(\ref{eq:su2bruder}) as a function of dissipation rate $\gamma_g/\gamma_d$ and the driving strength $\varepsilon$ in Fig.~\ref{fig:comparison}(c).
The synchronization blockade can be observed when $\gamma_d = \gamma_g$.
In Fig.~\ref{fig:comparison}(d) we plot $\Omega_{l_1}(\rho)$ for the same system assuming a diagonal limit cycle state.
The absence of synchronization blockade is further evidence that the set of limit cycle states $\Sigma$ must be expanded to include non-diagonal states satisfying the synchronization blockade condition in Eq.~(\ref{eq:blockade_cond}) if the system follows a proper subgroup $\mathcal{H}$.


\bigskip

\section{Discussion}
\label{sec:discussion}

We have presented an approach to study quantum synchronization blockade based on the symmetries of the physical system.
This allowed us to derive a general condition for the energy-level structure that the physical system must follow in order for quantum synchronization blockade to be possible.
In particular, we have shown that an $N$-level system whose dynamics is described by the $SU(N)$ algebra cannot support synchronization blockade.
Our approach is system-agnostic and can be readily applied to a variety of systems presented in the literature \cite{roulet2018synchronizing,jaseem2020quantum} as well as new ones with more complicated structure.

Understanding the conditions under which quantum synchronization blockade occurs is important from the perspective of measures of quantum synchronization.
Recently introduced approach of quantifying synchronization in quantum systems relies on identifying the structure of a suitable limit cycle state \cite{jaseem2020generalized}.
Our results show that systems that cannot support quantum synchronization blockade must have diagonal limit cycle states.
This result greatly simplifies the task of quantifying synchronization in quantum systems incapable of synchronization blockade.
For such systems, a simple $l_1$-norm of coherence is a suitable measure of quantum synchronization.
On the other hand, the $l_1$-norm is unsuitable for systems that support synchronization blockade and one must use partially coherent limit cycle states in order to quantify synchronization. This is accounted for in quantum information theoretic measures by a suitable choice of the limit-cycle state.

Synchronization and its disruption in classical dynamical systems have a variety of important applications including the treatment of functional brain disorders such as epilepsy \cite{jiruska2013synchronization}, design of anti-pacemakers for the destruction of pathological synchronization of a population of interacting oscillators \cite{kiss2007engineering}, or estimation of structural damping needed to stabilize crowded footbridges \cite{strogatz2005crowd}.
Apart from its fundamental importance, quantum synchronization is emerging as a phenomenon with interesting applications, particularly in the context of quantum engines \cite{jaseem2020quantum,solanki2022role}.
Quantum synchronization blockade is a genuinely quantum phenomenon leading to disruption of synchronization in quantum systems.
Therefore it is an exciting open question of how synchronization blockade may be applied in the broader context of quantum technologies.

\bibliographystyle{apsrev4-1}

%

\newpage

\appendix

\section{Contribution of Diagonal Elements to Synchronization Measure}\label{appendix1}
The completeness relation of a coherent state is given by
\begin{align}
    \mathcal{N}\int d\Omega_\phi\int d\Omega_\theta \ket{\alpha}\bra{\alpha}&=\mathbf{1},
\end{align}
where $d\Omega_\theta$ and $d\Omega_\phi$ are the contributions of the population degrees of freedom and free phases to the Haar measure ($d\Omega=d\Omega_\theta d\Omega_\phi$) respectively. The element-wise expression for the above equation can be written as
\begin{align}
 \mathcal{N}\int d\Omega_\phi \int d\Omega_\theta \alpha_j\alpha_k^*&=\delta_{jk},
 \label{eq:coherentstate}
\end{align}
such that $\mathcal{N}\int d\Omega_\phi \int d\Omega_\theta |\alpha_j|^2=1 \; \forall j
\leq N $ for $SU(N)$.
The term $|\alpha_i|^2$ is independent of free phases $\{\phi_j\}$, so Eq.~(\ref{eq:coherentstate}) can be simplified as follow
\begin{align}
    \mathcal{N}(2\pi)^{N-1} \int d\Omega_\theta |\alpha_j|^2=1,
    \label{eq:gen_integration1}
\end{align}
where the $(2\pi)^{N-1}$ comes from the integration over $N-1$ free phases. 
The above equation can be rearranged as follows,
\begin{align}
    \mathcal{N} \int d\Omega_\theta |\alpha_j|^2&=\left(\dfrac{1}{2\pi}\right)^{N-1}.\label{eq:gen_integration2}
\end{align}
Following from Eq.~(\ref{eq:final_gen_sync_meas}) of the main text, the contribution of the diagonal terms of the density matrix in the synchronization measure is given by
\begin{align}
    \sum_j \mathcal{N}R_{jj}z_{jj} &= \sum_{j} \mathcal{N}\int d\Omega_\theta |\alpha_j|^2 R_{jj},\nonumber\\
    &= \sum_{j}R_{jj}\left(\mathcal{N} \int d\Omega_\theta |\alpha_j|^2\right).
\end{align}
Using Eq.~(\ref{eq:gen_integration2}) and the unit trace property of the density matrix ($\sum_{j}R_{jj}=1$) in the above equation, we obtain
\begin{equation}
    \sum_j \mathcal{N}R_{jj}z_{jj} = \left(\dfrac{1}{2\pi}\right)^{N-1},
\end{equation}
which is the contribution of diagonal elements in the synchronization measure.

\section{Proof of Linear Independence}\label{appendix2}
In this section, we will show that for full $SU(N)$ systems, the synchronization measure is a linear combination of independent terms and vanishes only for diagonal steady states. 
Let us assume that we have a general $SU(N)$ system with $N$ free phases $\{\phi_1,\phi_2,\ldots,\phi_N\}$ (including the global phase) \cite{nemoto2000generalized}.
The synchronization blockade given by Eq.~(\ref{eq:blockade_cond}) can be simplified as follows,
\begin{align}
    \sum_{j<k}^Na_{jk}\cos(\phi_j-\phi_k-\chi_{jk})=0,
    \label{eq:new_block_cond}
\end{align}
We use the following notation, $C_{\alpha} \equiv \cos\alpha$ and $S_{\alpha} \equiv \sin\alpha$, in the rest of this section. 

Using the identity $\cos(\phi_j-\phi_k-\chi_{jk}) = C_{\phi_j}C_{\phi_k}C_{\chi_{jk}} + S_{\phi_j}S_{\phi_k}C_{\chi_{jk}} + S_{\phi_j}C_{\phi_k}S_{\chi_{jk}}-C_{\phi_j}S_{\phi_k}S_{\chi_{jk}}$ we can rewrite Eq.~(\ref{eq:new_block_cond}) as follows,
\begin{align}
    \sum_{j<k}a_{jk}C_{\phi_j}C_{\phi_k}C_{\chi_{jk}}&=\sum_{j<k}a_{jk}\Big(C_{\phi_j}S_{\phi_k}S_{\chi_{jk}}-S_{\phi_j}C_{\phi_k}S_{\chi_{jk}}\nonumber\\
    &-S_{\phi_j}S_{\phi_k}C_{\chi_{jk}}\Big).
\end{align}
Since the synchronization blockade condition is independent of free phases, we will use different values of the $\phi_i's$ to obtain the coefficients $a_{jk}$.
To begin with, we consider $\phi_j=0\,\,\forall\,\, j,$ which results in
\begin{align}
    \sum_{j<k}a_{jk}C_{\chi_{jk}}=0.
    \label{eq:coeff}
\end{align}
We now consider a different choice of parameters where $N-1$ phases are $\pi/2$ and one phase is zero such that $\phi_j=\pi/2\,\,\forall\,\, j\neq l$ and $\phi_l=0$. For the given choice of parameters, the Eq.~(\ref{eq:new_block_cond}) reduces to the following form
\begin{align}
    \sum_{k>l}a_{lk}S_{\chi_{lk}}-\sum_{j<l}a_{jl}S_{\chi_{jl}}&-\sum_{j<k; j,k\neq l}a_{jk}C_{\chi_{jk}}=0.
\end{align}
After adding and subtracting the terms $\sum_{j<l}a_{jl}C_{\chi_{jl}}+\sum_{k>l}a_{lk}C_{\chi_{lk}}$ from the left side of the above equation, it can be written as 
\begin{align}
    \sum_{k>l}a_{lk}S_{\chi_{lk}}-\sum_{j<l}a_{jl}S_{\chi_{jl}}&-\sum_{j<k}a_{jk}C_{\chi_{jk}}+\sum_{j<l}a_{jl}C_{\chi_{jl}}\nonumber\\
    &+\sum_{k>l}a_{lk}C_{\chi_{lk}}=0,\label{eq:B5}
\end{align}
where $\sum_{j<k}a_{jk}C_{\chi_{jk}}=  \sum_{j<l}a_{jl}C_{\chi_{jl}} + \sum_{k>l}a_{lk}C_{\chi_{lk}} + \sum_{j<k; j,k\neq l}a_{jk}C_{\chi_{jk}}$.
Equation~(\ref{eq:B5}) can be further simplified to read
\begin{align}
        \sum_{k>l}a_{lk}(S_{\chi_{lk}}+C_{\chi_{lk}})+\sum_{j<l}a_{jl}(S_{\chi_{lj}}+C_{\chi_{jl}})=0.\label{eq:B6}
\end{align}

For our next choice of parameters, we consider that $N-2$ phases are $\pi/2$ and  the rest of the two are zero such that $\phi_j=\pi/2\,\,\forall \,\,j\neq \{l,m\}$ and $ \phi_l=\phi_m=0$ where $l<m$. Therefore  Eq.~(\ref{eq:new_block_cond}) reduces to the following form
\begin{align}
    a_{lm}C_{\chi_{lm}}&=\sum_{k>l}a_{lk}S_{\chi_{lk}}-a_{lm}S_{\chi_{lm}}-\sum_{j<l}a_{jl}S_{\chi_{jl}}\nonumber\\
    &-\sum_{j<k}a_{jk}C_{\chi_{jk}}+\sum_{j<l}a_{jl}C_{\chi_{jl}}+\sum_{k>l}a_{lk}C_{\chi_{lk}}\nonumber\\
    &+\sum_{k>m}a_{mk}S_{\chi_{mk}}-\sum_{j<m}a_{jm}S_{\chi_{jm}}+a_{lm}S_{\chi_{lm}}\nonumber\\
    &-\sum_{j<k}a_{jk}C_{\chi_{jk}} +\sum_{j<m}a_{jm}C_{\chi_{jm}}+\sum_{k>m}a_{mk}C_{\chi_{mk}}\nonumber\\
    &-a_{lm}C_{\chi_{lm}}.
\end{align}
Using Eq.~(\ref{eq:coeff}), the above expression can be further simplified as 
\begin{align}
    2a_{lm}&C_{\chi_{lm}}=\sum_{k>l}a_{lk}(S_{\chi_{lk}}+C_{\chi_{lk}})+\sum_{j<l}a_{jl}(S_{\chi_{lj}}+C_{\chi_{jl}})\nonumber\\
    &+\sum_{k>m}a_{mk}(S_{\chi_{mk}}+C_{\chi_{mk}})+\sum_{j<m}a_{jm}(S_{\chi_{mj}}+C_{\chi_{jm}}).\label{eq:B8}
\end{align}
which leads to $a_{lm}=0 ~ \forall\, l,m $ following from Eq.~(\ref{eq:B6}). 
This shows that for an $N$-level system which is also described by the $SU(N)$ algebra, the synchronization measure vanishes only for diagonal limit cycle states since all the coherences ($a_{lm}$) are zero.
In other words, synchronization blockade is not observed in such systems. 

\section{Sufficient Condition for the Absence of Blockade}\label{appendix3}
In this section, we prove that a general $N$ level system having distinct eigenvalue separations will follow $\mathfrak{su}(N) $ algebra provided that all nearest-level transitions are allowed among $N$ levels.
To describe $N$ distinct energy levels, we require a set of $N-1$ traceless diagonal matrices $\{\lambda_j\}$ where $\lambda_j=\vert j \rangle \langle j \vert -\vert j+1 \rangle \langle j+1 \vert$.
The nearest-level transitions can be given by a set of $N-1$ traceless off-diagonal Hermitian matrices $\{\beta^R_j\}$ where $\beta^R_j=\vert j \rangle \langle j+1 \vert +\vert j+1 \rangle \langle j \vert$.
The given matrices $\{\mathcal{O}_j\}$ must follow the 
algebra defined as follows
\begin{align}
    [\mathcal{O}_j,\mathcal{O}_k]&=i\sum_{l}f_{jkl}\mathcal{O}_l, \\
    \{\mathcal{O}_j,\mathcal{O}_k\}&=\frac{1}{N}\delta_{jk}I+\sum_{l}d_{jkl}\mathcal{O}_l.
\end{align}
Commutation relationship between elements of $\{\lambda_j\}$ and $\{\beta_j^R\}$  generates additional set of $N-1$ matrices $\{\beta^C_j\}$ where $\beta^C_j=i(\vert j \rangle \langle j+1 \vert -\vert j+1 \rangle \langle j \vert)$.
To complete the group, the elements of $\{\beta^{R,C}_j\}$ must also follow the 
algebra discussed above giving rise to an additional set of $2(N-2)$ matrices $\{\gamma^{R,C}_j\}$ such that $\gamma_j^R=\vert j \rangle \langle j+2 \vert + \vert j+2 \rangle \langle j \vert$ and $\gamma_j^C=i\vert j \rangle \langle j+2 \vert -i \vert j+2 \rangle \langle j \vert$.
Following this, the given 
algebra will close for $ (N-1) + 2 [ (N-1)+(N-2)+\ldots+1 ] = N^2-1$ number of matrices which is the required number of matrices to complete the 
$\mathfrak{su}(N)$ algebra.
Therefore a system with $N$ non-degenerate eigenvalues separations having interaction between all nearest neighbor levels  will follow $SU(N)$ group.

Now let us consider that one of the levels is not interacting with others.
We still require $N-1$ diagonal traceless matrices $\{ \lambda_j^{'} \} $ to specify different energy level spacings. Since one of the levels is not interacting with others, we only require $2(N-2)$ off-diagonal matrices $\{\beta_j^{ '(R,C) } \} $ to model the nearest-level interactions.
This leads to $(N-1) + 2 [ (N-2)+(N-3)+\ldots+1 ] = (N-1)^2$ number of matrices required to close the algebra.
The given number of matrices is equivalent to the number of matrices required to specify the $\mathfrak{su}(N-1)$ 
algebra plus one more matrix to specify the energy eigenvalue of the isolated level following $U(1)$ group.

\end{document}